\documentclass[twocolumn,showpacs,amsmath,aps,pra,superscriptaddress]{revtex4}

\usepackage{graphicx}
\usepackage{dcolumn}
\usepackage{bm}

\newcommand{\ket}[1]{\left| {#1} \right\rangle}

\begin{document}

\title{Generation of polarization entanglement from spatially-correlated photons \\
in spontaneous parametric down-conversion} 

\author{Ryosuke Shimizu}
\affiliation{PRESTO, Japan Science and Technology Agency, 4-1-8 Honcho, Kawaguchi 332-0012, Japan}

\author{Takashi Yamaguchi}
\affiliation{Research Institute of Electrical Communication, Tohoku University, Sendai 980-8577, Japan}

\author{Yasuyoshi Mitsumori}
\affiliation{Research Institute of Electrical Communication, Tohoku University, Sendai 980-8577, Japan}
\affiliation{CREST, Japan Science and Technology Agency, 4-1-8 Honcho, Kawaguchi 332-0012, Japan}

\author{Hideo Kosaka}
\affiliation{Research Institute of Electrical Communication, Tohoku University, Sendai 980-8577, Japan}
\affiliation{CREST, Japan Science and Technology Agency, 4-1-8 Honcho, Kawaguchi 332-0012, Japan}

\author{Keiichi Edamatsu}
\affiliation{Research Institute of Electrical Communication, Tohoku University, Sendai 980-8577, Japan}
\affiliation{CREST, Japan Science and Technology Agency, 4-1-8 Honcho, Kawaguchi 332-0012, Japan}

\date{\hspace*{3cm}}

\begin{abstract}
We propose a novel scheme to generate polarization entanglement from spatially-correlated photon pairs. We experimentally realized a scheme by means of a spatial correlation effect in a spontaneous parametric down-conversion and a modified Michelson interferometer. The scheme we propose in this paper can be interpreted as a conversion process from spatial correlation to polarization entanglement.
\end{abstract}


\maketitle

Polarization entanglement of photons provides a superior environment in which to demonstrate quantum information and communication protocols in quantum mechanical 2-level states (qubits), e.g., quantum teleportation \cite{Bouwmeester97} and quantum cryptography \cite{Jennewein00}.
Thus far, several methods using spontaneous parametric down-conversion have been developed to generate a polarization-entangled state: one using a nonlinear crystal with type-II phase matching conditions \cite{Kwiat95}, another using two type-I crystals oriented at $90^\circ$ with respect to each other \cite{Kwiat99}, and a third using nonlinear crystals with an interferometer \cite{Branning00,Kim01,Shi04,T_Kim06}.
The phase-matching condition in a spontaneous parametric down-conversion causes strong correlations between the constituent photons with respect not only to polarization but also to other degrees of freedom, e.g., energy or momentum.
Recently, these degrees of freedom of photons have attracted much attention in the performance of quantum information protocols in a larger dimensional Hilbert space. 
For instance, such multidimensional quantum states (qudits) are expected to improve security for quantum key distribution \cite{Cerf02} and robustness against noise \cite{Collins02}.
Although there are some experimental demonstrations for creating entangled qudits using angular momentum \cite{Mair01} and time binning \cite{Riedmatten02}, the use of a transverse spatial degree of freedom provides a simple method for investigating entangled qudit states \cite{Neves05,Hale05}. 
In addition, entanglement in two or more degrees of freedom of photons has also been reported.
In particular, the state of being simultaneously entangled in multiple degrees of freedom, a so-called hyperentangled state, has been realized and characterized \cite{Barreiro05,Barbieri05}.
Actually, the polarization-spatial, hyperentangled state has been used to demonstrate a direct entanglement measure \cite{Walborn06}. 

In this article, we propose a novel approach to generating polarization-entangled photon pairs and demonstrate its experimental verification.
As described below, our scheme is interpreted as converting entanglement from spatial degrees of freedom to polarization ones. 

\begin{figure}
\centerline{\includegraphics[width=8cm]{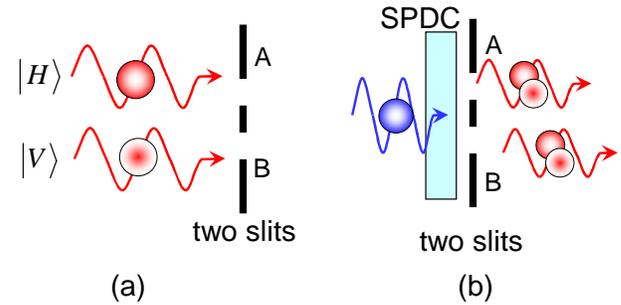}}
\caption{\label{fig1}(Color online) Situations proposed in our scheme.}
\end{figure}

To see the essence of our proposal, let us consider the situation in which two coherent, orthogonally-polarized photons pass through an object having two slits [see Fig.~\ref{fig1} (a)].
Together with the polarization and spatial degrees of freedom, we can consider the following symmetrical two-photon state:
\begin{multline}
\ket{\psi} =  \frac{1}{2\sqrt{2}} \left(  \ket{H,A}\ket{V,A} + \ket{H,B}\ket{V,B} \right. \\
           \left. + \ket{H,A}\ket{V,B} + \ket{H,B}\ket{V,A} \right) + \textrm{t.t.} \label{eq1}
\end{multline}
Here $H$ ($V$) represents the horizontal (vertical) polarization state of a photon, $A$ ($B$) the spatial mode as determined by the slit $A$ ($B$), and t.t. means the transposed terms in which the first and second photons are transposed. 
In this situation, we assume that the first and second photons are indistinguishable from one another except for the polarization or spatial degrees of freedom.
Thus the transpose operation is identical for both photons in the two-photon state.
Hereafter, we omit the transposed terms for simplicity.

A signal and its conjugate idler photons are generated in almost the same position in a nonlinear crystal as explained by Fourier-optical analysis for the two-photon state generated by spontaneous parametric down-conversion \cite{Shimizu03,Shimizu06,Angelo01}.
Thus, if frequency-degenerate photon pairs are collinearly generated by a type-II phase matching condition and two slits are placed just after the crystal [see Fig.~\ref{fig1} (b)], we can prepare the following two-photon state as a result of a spatial correlation effect:
\begin{equation}
\ket{\phi} = \frac{1}{\sqrt{2}} \left( \ket{H,A}\ket{V,A} + \ket{H,B}\ket{V,B} \right). \label{eq2}
\end{equation}
Such an effect has attracted great interest because of its possible application to a novel imaging technology that is called ``quantum imaging'' \cite{Lugiato02}.
The two-photon state in Eq.~(\ref{eq2}) is a spatially-entangled state in which both photons pass together through either of the slits; the photons are at this time not entangled in polarization degrees of freedom. 
If we exchange the spatial mode with each other for one of the polarization mode, an action which corresponds to a controlled-NOT (CNOT) operation on the spatial part depending on the polarization part \cite{Fiorentino04_A}, the two-photon state in Eq.~(\ref{eq2}) is converted to the following state:
\begin{equation}
\ket{\phi'} = \frac{1}{\sqrt{2}} \left( \ket{H,A}\ket{V,B} + \ket{H,B}\ket{V,A} \right). \label{eq3}
\end{equation}
Since these two-photon states have identical forms in the transpose operation, the polarization state of Eq.~(\ref{eq3}) is rewritten with respect to the spatial mode as follows:
\begin{equation}
\ket{\phi'} = \frac{1}{\sqrt{2}}\left( \ket{H}_{A} \ket{V}_{B} + \ket{V}_{A} \ket{H}_{B} \right). \label{eq3a}
\end{equation}
This is the standard notation of a polarization-entangled state.
The scheme we proposed here can be understood as an entanglement conversion process from spatial degrees of freedom to polarization ones.

\begin{figure}
\centerline{\includegraphics[width=8cm]{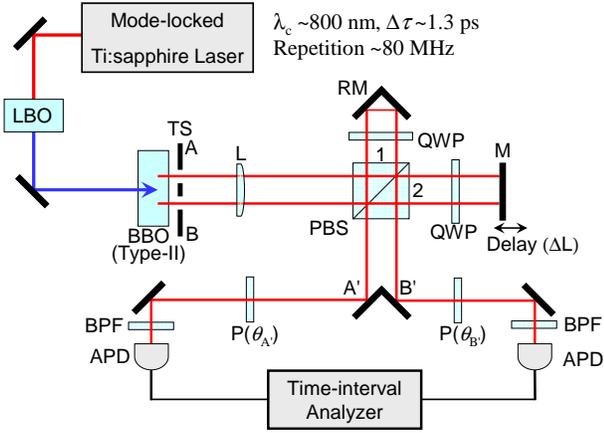}}
\caption{\label{fig2}(Color online) Schematic drawing of our experimental setup. LBO, Lithium Triborate crystal for frequency doubling; TS, two slits; L, collimating lens; PBS, polarizing beam splitter; QWP, quarter-wave plate; P, polarization analyzer; BPF, 3-nm band pass filter centered at 800 nm; APD, avalanche photo diode. QWPs are used to change polarization from $H (V)$ to $V (H)$.}
\end{figure}

A schematic diagram of the setup used to realize our proposal described above is depicted in Fig.~\ref{fig2}.
A well-collimated, frequency-doubled, mode-locked Ti:sapphire laser operating at 400 nm, whose average power is 110 mW and repetition rate is 80 MHz, pumps a $\beta$-barium borate (BBO) crystal.
Frequency-degenerate photon pairs are collinearly generated according to type-II phase matching conditions.
Putting two slits (each slit width: 150 $\mu$m; interval: 450 $\mu$m) just after the nonlinear crystal, we can produce the two-photon state described in Eq.~(\ref{eq2}).
In this condition, the pump beam diameter on the two slits is approximately 1.7 mm.
After passing through a cylindrical lens (focal length: 50 mm), the photon pairs are fed into a modified polarization Michelson interferometer that consists of a polarization beam splitter (PBS), a roof mirror (RM) and a plane mirror (M).
In the PBS, the photon pairs are divided into two arms of the interferometer, depending on their polarization. In arm 1, where the roof mirror is placed, the path mode $A$ ($B$) of a $V$-polarized photon changes into $B'$ ($A'$), while the path mode of a $H$-polarized photon in arm 2 does not change.
Thus, after the recombination in the PBS, the output state from through the interferometer produces the polarization-entangled state described in Eq.~(\ref{eq3}).
The output photons from the interferometer were detected with avalanche photodiodes (EG\&G SPCM AQ161) through a pair of polarization analyzers consisting of a half-wave plate (HWP) and a polarizing beam splitter, and through 3-nm interference filters centered at 800 nm.
We recorded the number of coincident events with a time-interval analyzer (EG\&G 9308).

\begin{figure}
\centerline{\includegraphics[width=8cm]{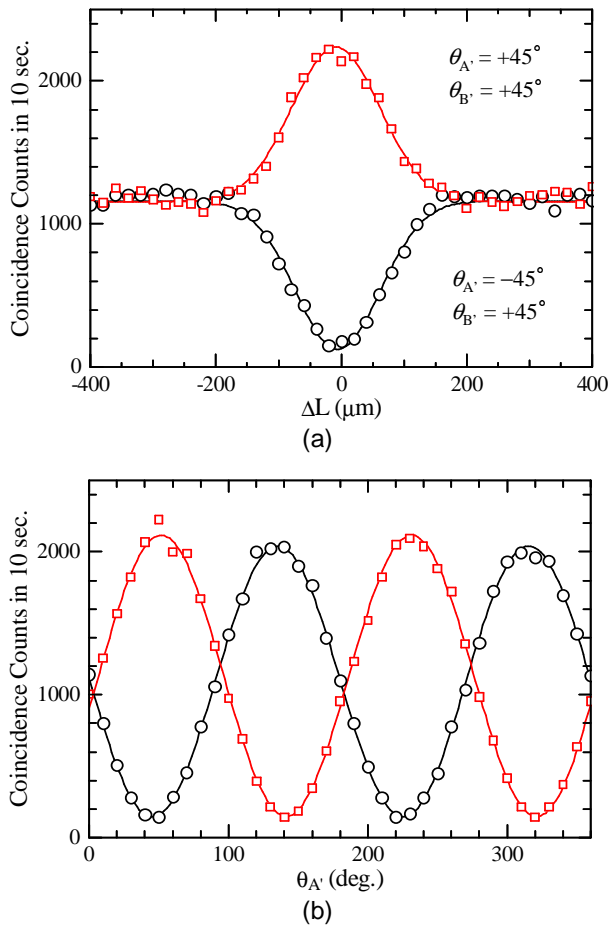}}
\caption{\label{fig3}(Color online) Experimental results (open circles and squares) and fitted curves of coincidence counts as a function of (a) the path-length difference ($\Delta L$) of the modified polarization Michelson interferometer and (b) the polarization analyzer angle ($\theta_{A'}$) in the path $A'$.}
\end{figure}

Figure~\ref{fig3}(a) shows the coincidence counts as a function of the path-length difference ($\Delta L$) of the interferometer.
In this measurement, we set the polarization analyzer angle in the path $A'$ ($\theta_{A'}$) to $+45^\circ$ (open squares) or $-45^\circ$ (open circles), while that in the path $B'$ ($\theta_{B'}$) was fixed at $+45^\circ$.
We observed a constructive quantum interference fringe with a visibility of 89 \% when $\theta_{A'}$ was set to $+45^\circ$ and a destructive interference fringe when $\theta_{A'}$ was set to $-45^\circ$. In addition, we performed polarization correlation measurements at $ \Delta L = 0\ \mu$m.
In these measurements, as shown in Fig.~\ref{fig3}(b), $\theta_{B'}$ was set to $+45^\circ$ (open squares) or $-45^\circ$ (open circles), and $\theta_{A'}$ was varied by rotating the HWP.
These resultant quantum interference fringes indicate that the output state forms a triplet state in polarization degrees of freedom: $\left( \ket{H}_{A'} \ket{V}_{B'} + \ket{V}_{A'} \ket{H}_{B'} \right)/\sqrt{2}$.
Thus, we have successfully demonstrated our proposal utilizing spatial correlation in a spontaneous parametric down-conversion.
The slight degradation of the quantum interference visibility observed in Fig.~\ref{fig3} was likely caused by a spatial phase disorder and mismatch of the transverse intensity profile between the spatial modes $A$ and $B$.
Controlling the phase relationship by a spatial phase modulator, we could obtain quantum interference with higher visibility.
In addition, while the inteference fringes in Fig.~\ref{fig3}(b) should be out of phase by 180 degrees with each other, in practice, they show the slightly asymmetric patterns.   
This discrepancy may come from the imbalance of the pump beam intensity between the slit $A$ and $B$, which results in the imbalance of the probability amplitude in Eq.~(\ref{eq3a}). 

\begin{figure}
\centerline{\includegraphics[width=8cm]{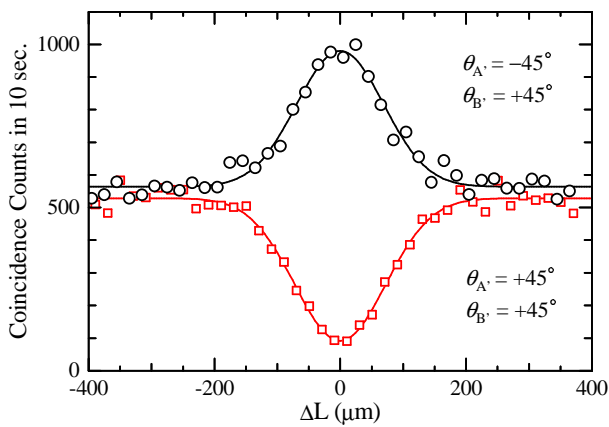}}
\caption{\label{fig4}(Color online) Experimental results (open circles and squares) and fitted curves of coincidence counts as a function of the path-length difference ($\Delta L$) of the modified polarization Michelson interferometer at particular angles of the mirror M.}
\end{figure}

Another remarkable feature of our polarization-entangled photon source is that a relative phase between the two terms in Eq.~(\ref{eq3}) is insensitive to the path-length difference $\Delta L$ unlike other polarization-entangled photon sources using a Mach-Zehnder interferometer \cite{Kim01} and a Michelson interferometer \cite{Branning00}.
On the other hand, the relative phase is sensitive to a tilt of the mirror M because it causes a path-length difference between the photon in the path mode $A'$ and that in $B'$ for $H$-polarization in arm 1.
As shown in Fig.~\ref{fig4} at particular angles of the mirror M, we observed constructive interference with the visibility of 82 \% when the polarization analyzer was set to  $A' = -45^\circ$ and  $B' = +45^\circ$ (open circles), and we observed destructive interference with settings of $ A' = +45^\circ$ and $B' = +45^\circ$ (open squares).
These observations stand in contrast to the two-photon interference fringes discussed in Fig.~\ref{fig3}.
These results indicate that with the same apparatus we can control the relative phase, without sacrificing the system stability, by tuning the angle of the mirror M and consequently produce a singlet state, $\left( \ket{H}_{A'} \ket{V}_{B'} - \ket{V}_{A'} \ket{H}_{B'} \right)/\sqrt{2}$.

Using this polarization-entangled photon source, we have tested the violation of Clauser-Horne-Shimony-Holt (CHSH) Bell inequality \cite{Clauser69}.
Following the method described in Ref. \cite{Aspect82}, we estimated the Bell parameter $S$ from coincidence counting measurements for 16 combinations of polarization analyzer settings $( \theta_{A'} = 0^\circ, 45^\circ, 90^\circ, 135^\circ;  \theta_{B'} = 22.5^\circ, 67.5^\circ, 112.5^\circ, 157.5^\circ)$.
In each measurement, we took coincidence counts for 10 s and obtained the $S$-value of $2.61 \pm 0.04$, which clearly indicates a violation of the classical limit $S = 2$.
This result shows that the output two-photon state in polarization degrees of freedom has a nonlocal correlation.

In summary, we have proposed and demonstrated a novel scheme for the generation of polarization entanglement utilizing a spatial correlation effect in spontaneous parametric down-conversion.
We have also shown a violation of Bell's inequality for the entangled photon source described here.
Our scheme is similar to the scheme proposed by Kim \textit{et al.} \cite{T_Kim06} at the fundamental level; the spatial entanglement is created with respect to the path mode of the Sagnac interferometer in Kim's scheme.
The brightness of our source is relatively lower than that of Kim's scheme, because the use of multiple slits to create the spatial modes results in waste of the pump power.  
However, our scheme has the advantage that we can easily extend the number of spatial modes.
By increasing the number of slits, we can extend the number of spatial modes so as to prepare the entanglement in a Hilbert space with larger dimensions.
In addition, our scheme has a potential to improve the brightness by adopting waveguide structure to create the spatial modes instead of the multiple slits.
The use of a quasi-phase-matched device with multiple waveguides may allow us to achieve construction of a high-flux entangeld photon source.

We are grateful to H. Ishihara for his valuable discussions.
This work was supported in part by Strategic Information and Communications R \& D Promotion Program (SCOPE) of the Ministry of Internal Affairs and Communications and by a Grant-in-Aid for Creative Scientific Research (17GS1204) of the Japan Society for the Promotion of Science.

\end{document}